# Two Ising-like magnetic excitations in a single-layer cuprate superconductor

Yuan Li[1,2], G. Yu[3], M.K. Chan[3], V. Balédent[4], Yangmu Li[3], N. Barišić[3,5], X. Zhao[3,6], K. Hradil[7,†], R.A. Mole[8,‡], Y. Sidis[4], P. Steffens[9], P. Bourges[4], M. Greven[3,*]

[1]Department of Physics, Stanford University, Stanford, California 94305, USA

[2]Max Planck Institute for Solid State Research, 70569 Stuttgart, Germany

[3]School of Physics and Astronomy, University of Minnesota, Minneapolis, Minnesota 55455, USA

[4]Laboratoire Léon Brillouin, CEA-CNRS, CEA-Saclay, 91191 Gif sur Yvette, France

[5]1. Physikalisches Institut, Universität Stuttgart, 70550 Stuttgart, Germany

[6]State Key Lab of Inorganic Synthesis and Preparative Chemistry, College of Chemistry, Jilin University, Changchun 130012, China

[7]Institut für Physikalische Chemie, Universität Göttingen, 37077 Göttingen, Germany

[8]Forschungsneutronenquelle Heinz Maier-Leibnitz, 85747 Garching, Germany

[9]Institut Laue Langevin, 38042 Grenoble CEDEX 9, France

† Present address: X-ray Center, Vienna University of Technology, 1060 Vienna, Austria

‡ Present address: The Bragg Institute, Australian Nuclear Science and Technology Organization, Locked Bag 2001, Kirrawee DC, NSW 2232, Australia

* To whom correspondence should be addressed: greven@physics.umn.edu

There exists increasing evidence that the phase diagram of the high-transition temperature ($T_c$) cuprate superconductors is controlled by a quantum critical point. One distinct theoretical proposal is that, with decreasing hole-carrier concentration, a transition occurs to an ordered state with two circulating orbital currents per $CuO_2$ square. Below the 'pseudogap' temperature $T^*$ ($T^* > T_c$), the theory predicts a discrete order parameter and two weakly-dispersive magnetic excitations in structurally simple compounds that should be measurable by neutron scattering. Indeed, novel magnetic order and one such excitation were recently observed. Here, we demonstrate for tetragonal $HgBa_2CuO_{4+\delta}$ the existence of a second excitation with local character, consistent with the theory. The excitations mix with conventional antiferromagnetic fluctuations, which points toward a unifying picture of magnetism in the cuprates that will likely require a multi-band description.

It is widely agreed that attaining a thorough understanding of the peculiar electronic and magnetic properties in the pseudogap regime of the cuprates would constitute a major leap toward solving the high-$T_c$ problem. A pivotal and intensely debated question has been whether this regime is a genuine new phase of matter and, if so, what symmetries are broken at $T^*$[1-4]. There is mounting evidence that $T^*$ indeed marks a transition into a novel electronic phase in which time-reversal symmetry is broken[5-10] and, in compounds with relatively high maximal transition temperatures ($T_{c,max} > 90$ K at the optimal hole concentration $p_{opt} \approx 16\%$ per planar Cu atom), translational symmetry is preserved[6-8,11].

Neutron scattering is a powerful probe of magnetic correlations and has shed much light on the high-$T_c$ problem. In the superconducting doping regime, magnetic neutron scattering experiments have been carried out mostly near the two-dimensional (2D) wave vector $\mathbf{q}_{AF}$ that characterizes the antiferromagnetic order of the undoped Mott-insulating parent compounds[12-20]. A spin-1 'resonance' excitation[13,15-17,21] is observed at $\mathbf{q}_{AF}$ in the superconducting state, between nearly temperature-independent spin fluctuations at higher energy and a magnetic gap at lower energy. This phenomenon has been regarded as indicative of a magnetic-fluctuation-driven superconducting mechanism[22,23]. On the other hand, recent measurements of the $T_{c,max} > 90$ K compounds $YBa_2Cu_3O_{6+\delta}$ (YBCO)[6] and $HgBa_2CuO_{4+\delta}$ (Hg1201)[7,8] revealed a novel kind of magnetic order (broken time-reversal symmetry) below $T^*$ that is characterized by the wave vector $q = 0$ (preserved lattice translational symmetry). The measurements were motivated by the distinct theoretical proposal that magnetism due to orbital charge currents (rather than local spin moments) lies at the heart of the cuprate phase diagram[1]. The subsequent discovery of a prominent magnetic excitation in Hg1201, which also appears below $T^*$ and is centered at $q = 0$, appears to be the first dynamic fingerprint of this pseudogap magnetism[24]. However, it has remained largely elusive if and how the antiferromagnetism and the pseudogap magnetism are related. Here we use inelastic neutron scattering to further determine the excitation spectrum associated with the latter. Our new results for Hg1201 reveal a second weakly-dispersive magnetic excitation branch, as predicted theoretically[25,26], as well as an intriguing mixing with the antiferromagnetic fluctuations near $\mathbf{q}_{AF}$ that is not yet captured theoretically.

Hg1201 possesses a simple tetragonal crystal structure, exhibits the highest value of $T_{c,max}$ ($\approx 96$ K) of all single-layer cuprates (one $CuO_2$ layer per primitive cell), and is thought to be relatively free of disorder effects[27,28]. Sizable crystals of Hg1201 have become available only in recent years[29] and enabled initial neutron scattering experiments[7,8,24,30,31]. Our underdoped ($T_c = 65$ K, $T^* \approx 330$ K, mass = 1.8 g; denoted UD65) and a nearly optimally doped ($T_c = 95$ K, $T^* \approx 210$ K, mass = 2.0 g; denoted OP95) samples[24] were measured with both spin-polarized and unpolarized neutrons. Scattering wave vectors are quoted as $\mathbf{Q} = H\mathbf{a}^* + K\mathbf{b}^* + L\mathbf{c}^* \equiv (H,K,L)$ in units of the reciprocal lattice vectors (r.l.u.), with typical room-temperature values $a^* = b^* = 1.614$ Å$^{-1}$ and $c^* = 0.657$ Å$^{-1}$. Further experimental details are provided in the Supplementary Information (SI).

We first provide evidence for magnetic excitations below $T^*$ from measurements with unpolarized neutrons. Figure 1a-c shows energy scans at various locations in the first 2D Brillouin zone. Since both nuclear and magnetic scattering contribute to the intensity, we use the intensity difference between the lowest temperature and a high temperature (close to $T^*$) to extract magnetic signals, based on the expectation that phonon intensity decreases upon cooling, whereas magnetic intensity increases. Especially for UD65, this method clearly reveals the presence of two weakly dispersive excitation branches throughout the entire Brillouin zone, with approximate energies of 38 and 54 meV (Fig. 1b). The branch near 54 meV was the subject of our previous study, and its magnetic origin was verified with spin-polarized neutrons[24]. The result in Fig. 1e further confirms this conclusion: apart from an enhancement near $\mathbf{q}_{AF}$ due to the presence of conventional antiferromagnetic fluctuations (Supplementary Figs. S1a and S2), the signal gradually decreases toward large in-plane momentum transfer, consistent with a magnetic origin. A phonon-based interpretation is further ruled out by the comparison between the scattering at $(0, 0, 4.6)$ and $(2, 2, 4.6)$, because the phonon dynamic structure factor at $(0, 0, 4.6)$ cannot be larger than at $(2, 2, 4.6)$, yet the intensity at the former position is clearly larger (Supplementary Fig. S3). A similar decrease of intensity with increasing $Q$ is found for the low-energy excitation branch in UD65 (Fig. 5a and Supplementary Fig. S6a), implying that it is also of magnetic origin. Figure 1d summarizes our results for the dispersion of the two branches along $[H, H]$.

Although the presence of a low-energy excitation is not as evident for OP95 as for UD65, there is a clear difference between the data in Figs. 1b and c: unlike for UD65, for OP95 there is no peak at ~ 38 meV, but instead a 'shoulder' near 31 meV. This is best seen in Fig. 2a by comparing the '4 K - 330 K' intensity difference for both samples, measured at (0, 0, 4.6) under nearly identical experimental condition. Given the rather small difference in oxygen concentration between OP95 and UD65 ($\Delta\delta \sim 0.03$, assuming each oxygen dopes two holes), the difference in the data is rather unlikely to be due to phonons and more naturally explained by a shift of the excitation from ~ 38 meV in UD65 to ~ 31 meV in OP95, reflecting a doping dependence of the underlying magnetism.

The presence of a magnetic signal at ~ 31 meV in OP95 is further supported by the data in Fig. 2c, which reveal that the intensities of the two excitations depend on the momentum transfer direction in a peculiar, opposite fashion. It was previously found that the high-energy excitation becomes indiscernible when $\mathbf{Q}$ is parallel to the $CuO_2$ planes (Supplementary Fig. S4 of ref. 24), which is confirmed in Fig. 2c. Conversely, although non-zero intensity is observed for the low-energy excitation for $\mathbf{Q}||c$, higher intensity is observed for $\mathbf{Q}||ab$ with both unpolarized (Fig. 2c) and polarized neutrons (Fig. 4b,c). The low-energy features at both $\mathbf{Q}$ positions are more clearly observed from the '4 K - 110 K' intensity difference (Fig. 2b), since the lower reference temperature improves the clarity of the result because variation in phonon scattering is kept to a minimum. The opposite momentum dependence of the intensities implies that the two excitation branches are associated with fluctuations in perpendicular directions, either purely in the magnetic degrees of freedom, or in conjunction with lattice vibrations. However, without an extensive study of the neutron spin-polarization dependence of the signal beyond the present work (especially of the low-energy branch with $\mathbf{Q}\|ab$, which would allow for a differentiation between magnetic fluctuations parallel and perpendicular to the copper-oxygen planes) a conclusive explanation of this phenomenon is unreachable. Here we simply regard it as empirical evidence that the two branches have the same physical origin. This is further evinced by the fact that the excitations exhibit similar intensity amplitudes (Figs. 1b-c and 2a) and temperature dependences (Fig. 3), with an onset temperature consistent with $T^*$ determined from resistivity and neutron diffraction[21]. No well-defined magnetic signal is observed in the raw data above $T^*$ (Fig. 1a) or in the intensity difference for temperatures above $T^*$ (Fig. 2d). Together with the fact that the excitations emanate from $q = 0$ (Figs. 1e and 5a), this demonstrates that they are associated with the $q = 0$ magnetic order.

We used spin-polarized neutrons (see SI for a detailed description of the method) to further verify the magnetic origin of the low-energy excitation branch. Such measurements are extremely difficult, not only because of the much reduced neutron flux, but also because a large part of background intensity arises from incoherent scattering and can not be suppressed further in spin-flip measurements. Moreover, imperfect shielding leads to additional (small) background intensity which is not negligible compared to the weak signal strength in the polarized measurements. Altogether this results in a much reduced signal, but not necessarily an improved signal-to-background ratio compared to unpolarized measurements, hence extremely long counting times are required (see Supplementary Fig. S4 for a comparison between polarized and unpolarized measurements).

In Fig. 4a-c, the intensity difference between low and high temperatures for OP95, measured in the spin-flip scattering geometry, shows a peak at ~ 31 meV near the 2D zone center (Fig. 4b) and also for $L = 0$ (Fig. 4c), consistent with the unpolarized results (Figs. 1c, 2a-c, 3b-c). Since no prominent nuclear scattering feature is observed in the non-spin-flip geometry (Fig. 4d,e), the experiment's flipping ratio of about 10 (which is high for inelastic scattering at these energies) ensures that the observed spin-flip signal is not due to polarization leakage. We note, on the other hand, that the data do not allow us to rule out a non-spin-flip contribution that is comparable in strength to the spin-flip signal. Hence it is not impossible that the excitations contain an admixture with lattice vibrations.

A more stringent test of magnetic scattering utilizes the polarization dependence of any genuine magnetic signal: spin-flip scattering probes magnetic fluctuations perpendicular to both the momentum transfer, $\mathbf{Q}$, and the spin polarization of the incident neutrons, $\mathbf{S}$. As a result, the magnetic signal is maximized when $\mathbf{S}$ is parallel to $\mathbf{Q}$ ($\mathbf{S}\|\mathbf{Q}$), whereas all other scattering processes are independent of the orientation of $\mathbf{S}$. The purely magnetic signal can be extracted by taking the intensity ($I$) combination: $2 \times I_{\mathbf{S}\|\mathbf{Q}} - I_{\mathbf{S}\perp\mathbf{Q}} - I_{\mathbf{S}\|\mathbf{Z}}$ (see SI for details), where $\mathbf{S}\perp\mathbf{Q}$ and $\mathbf{S}\|\mathbf{Z}$ denote the two geometries in which $\mathbf{S}$ is perpendicular to $\mathbf{Q}$, horizontal and vertical, respectively. Based on Fig. 4a,f, Fig. 4g therefore demonstrates the presence of magnetic intensity centered at ~ 30 meV (OP95) and ~ 37 meV (UD65), in excellent agreement with the unpolarized-neutron data in Figs. 4h, 1b&c, and 2a-c. While the individual errors in Fig. 4g are relatively large, statistical analysis shows it to be a very robust result that the excitation (established to be present with unpolarized neutrons) is indeed predominantly magnetic (see SI).

Our results provide valuable insight into the fundamental properties of the pseudogap magnetism. The very weak dispersion of about 5% (Fig. 1d) and the absence of a Goldstone mode dispersing to zero energy at the ordering wave vector $q = 0$ imply that the order parameter has discrete symmetry. The dispersion is even weaker than that of the classic local-moment Ising-like antiferromagnet $Rb_2CoF_4$, in which the spin excitations disperse by about 20%[32]. Contrary to this model magnet, we observe two excitation branches rather than one. Together, these results suggest the presence of multiple scattering centers per $CuO_2$ square (or $CuO_6$ octahedron) and the need for a multi-band rather than a single-band theoretical description. The orbital-current theory, which is based on a multi-band Hamiltonian and makes the non-trivial prediction of two magnetic collective excitations in a single-layer system measurable via neutron scattering, appears to be able to explain our findings[25,26,33]. In this model, the weak dispersion is a direct consequence of the underlying discrete order parameter, whereas the non-degeneracy of the excitations has been suggested to be due to the nature of the ground and excitation states, which are quantum superpositions of four 'classical' degenerate orbital-current patterns[26,33]. This superposition has also been proposed to account for the peculiar experimental result that the magnetic moment direction is neither perpendicular nor parallel to the $CuO_2$ layers[6-8,34]. On general grounds, mode softening is expected at high temperature and upon approaching the quantum critical point. The former is not observed in our experiment and would require high-statistics energy scans at temperatures just below $T^*$. However, with increasing doping, we observe a clear softening of the low-energy branch.

Our results are consistent with the orbital-current theory. We note though that a distinctly different possibility consistent with the very weak dispersion is that the excitations are related to intrinsic inhomogeneity in the local electronic environment[35,36]. It has been proposed that such inhomogeneity can give rise to local 'edge modes' that are partially magnetic[35].

Our data reveal an intriguing connection between the pseudogap excitations and the conventional antiferromagnetic fluctuations at $\mathbf{q}_{AF}$. Initial evidence comes from the prior observation for OP95[24] that the resonance occurs at an energy that is indistinguishable from that of the high-energy pseudogap excitation, which is confirmed with improved precision in Supplementary Figs. S1a and S2. A local intensity maximum at $\mathbf{q}_{AF}$ is also found for the low-energy excitation in OP95 (Supplementary Fig. S1b), but the relatively weak signal does not allow for a detailed study. Even though there exists no clear resonance (distinct intensity change) across $T_c$ in UD65, we observe an enhanced response at $\mathbf{q}_{AF}$ at 39 meV, the energy of the pseudogap excitation (Fig. 5a). Figure 5b provides a detailed view of the response near $\mathbf{q}_{AF}$ along $\mathbf{a}^*$. For YBCO, this momentum direction is optimal for observing the 'hourglass' dispersion of the antiferromagnetic fluctuations in the superconducting state[37]. Indeed, we find initial evidence for a similar concave dispersion near $\mathbf{q}_{AF}$ in Hg1201, with a maximum energy that is indistinguishable from that of the lower pseudogap excitation. The signal amplitudes of the antiferromagnetic fluctuations, determined from momentum scans (which are insensitive to the pseudogap excitations because of the weak dispersion), are comparable to those of the pseudogap excitations in Hg1201, and to those of antiferromagnetic fluctuations in other cuprates (e.g., YBCO). Moreover, the signal that peaks at $\mathbf{q}_{AF}$ exhibits two local maxima at approximately the same energies as the pseudogap excitations (Supplementary Fig. S6, summarized in Fig. 5c). Evidently, the two types of excitations mix, even though they appear to have rather different physical origins: while the fluctuations near $\mathbf{q}_{AF}$ are generally thought to arise from copper spin moments, the weakly-dispersive pseudogap excitations appear to require the explicit consideration of oxygen orbitals and are best explained by the orbital-current theory.

Understanding the confluence of the two types of magnetic excitations will require a unifying theoretical approach. In the orbital-current theory, the superconducting pairing is the result of quantum critical fluctuations associated with the discrete pseudogap order parameter[38], and antiferromagnetic correlations have not yet been included. On the other hand, theories in which the pairing is mediated by antiferromagnetic fluctuations[39-41] have generally ignored the possibility that the pseudogap regime is a genuine new phase. Since the superconductivity is an instability of the peculiar 'normal' state, our results imply that even if antiferromagnetic fluctuations play a role in bringing about superconductivity in the cuprates, they must not be thought of as mere remnants of the Mott-insulating state, but rather as fundamentally modified by the pseudogap state that is characterized by weakly-dispersive excitations. In fact, the size of the superconducting gap ($\Delta$) appears to be defined already at $T^*$: the magnetic resonance energy in unconventional superconductors has been shown to be universally proportional to $\Delta$[21] and, in the model compound Hg1201, the resonance occurs at the same energy as the high-energy pseudogap excitation.

Bearing in mind that the pseudogap excitations and the antiferromagnetic fluctuations in Hg1201 occur at the same energy, we note that there might exist a correspondence between magnetic energy scales of single-layer Hg1201 and double-layer YBCO, two cuprates with similar values of $T_{c,max}$ and $\Delta$, and with well-defined resonances at $\mathbf{q}_{AF}$ near optimal doping[13,15-17,30]. In YBCO, the presence of two resonances in the 30-60 meV range has been interpreted as due to the interaction between the two adjacent $CuO_2$ layers in the same primitive cell[42]. Surprisingly, we find that the energies of the pseudogap excitations in UD65 Hg1201 ($39 \pm 2$ meV and $56 \pm 2$ meV at $\mathbf{q}_{AF}$) are equal within the error to those of the odd ($\approx$ 37 meV) and even ($\approx$ 55 meV) parity resonances in YBCO with a similar $T_c$ ($\approx$ 63 K)[43]. This observation also holds for the high-energy mode of OP95 Hg1201 ($55 \pm 2$ meV at $\mathbf{q}_{AF}$), but not for the corresponding low-energy mode ($32 \pm 3$ meV at $\mathbf{q}_{AF}$): in nearly optimally-doped YBCO ($T_c \approx 89$ K), the two resonance energies are about 53 and 41 meV[42].

The pseudogap excitations should be most easily discernable in compounds in which the $q = 0$ order is prominent, and so far they have been reported only for Hg1201. The well-studied single-layer materials (La,Nd,Sr,Ba)$_2$CuO$_4$ possess a relatively low $T_{c,max}$ of about 40 K and have long been known to exhibit an instability toward broken translational symmetry (spin/charge 'stripe' order) well below $T*$[18]. The lack of evidence of pseudogap excitations in these compounds likely results from a competition between the two types of order[34].

On the other hand, it should be possible to observe the pseudogap excitations in YBCO ($T_{c,max} \approx 93$ K). At low doping, near the onset of superconductivity, neutron diffraction measurements have revealed a quasi-elastic signal consistent with a transition to long-range spin-density-wave order as $T \rightarrow 0$[20]. The spin-density-wave and $q = 0$ orders are associated with very different wave vectors and appear to compete in the deeply underdoped regime ($p < 0.09$)[44], whereas the $q = 0$ order is found to dominate at higher doping[6], where the pseudogap excitations are most likely to be found. Material-specific differences, such as the more complicated double-layer structure of YBCO, can be expected to cause variations in the number of pseudogap excitations and in their strength relative to antiferromagnetic fluctuations. Analogous to the situation for single-layer LSCO and Hg1201, the pseudogap magnetism in the double-layer compounds might eventually be most clearly revealed in HgBa$_2$CaCu$_2$O$_{8+\delta}$ ($T_{c,max} \approx 124$ K[45], the highest value for all double-layer compounds) once sizable single crystals become available.

**Acknowledgements** We thank C. M. Varma, A. R. Bishop and B. Keimer for stimulating discussions. The research project was supported by the US Department of Energy, office of Basic Energy Sciences. X.Z. acknowledges support by the National Natural Science Foundation, China. Y. Li acknowledges support from the Alexander von Humboldt Foundation.

**Author Contributions** M.G., P.B., and Y.L. planned the project. Y.L., G.Y., M.K.C.,

V.B., and Y.-M.L. performed the neutron scattering experiments. Y.L., N.B., and X.Z. characterized and prepared the samples. P.S., R.A.M., K.H., Y.S. and P.B. were local contacts for the neutron scattering experiments. Y.L. and M.G. analyzed the data and wrote the manuscript with input from all coauthors.

**Author Information** Correspondence and requests for materials should be addressed to M.G. (greven@physics.umn.edu).

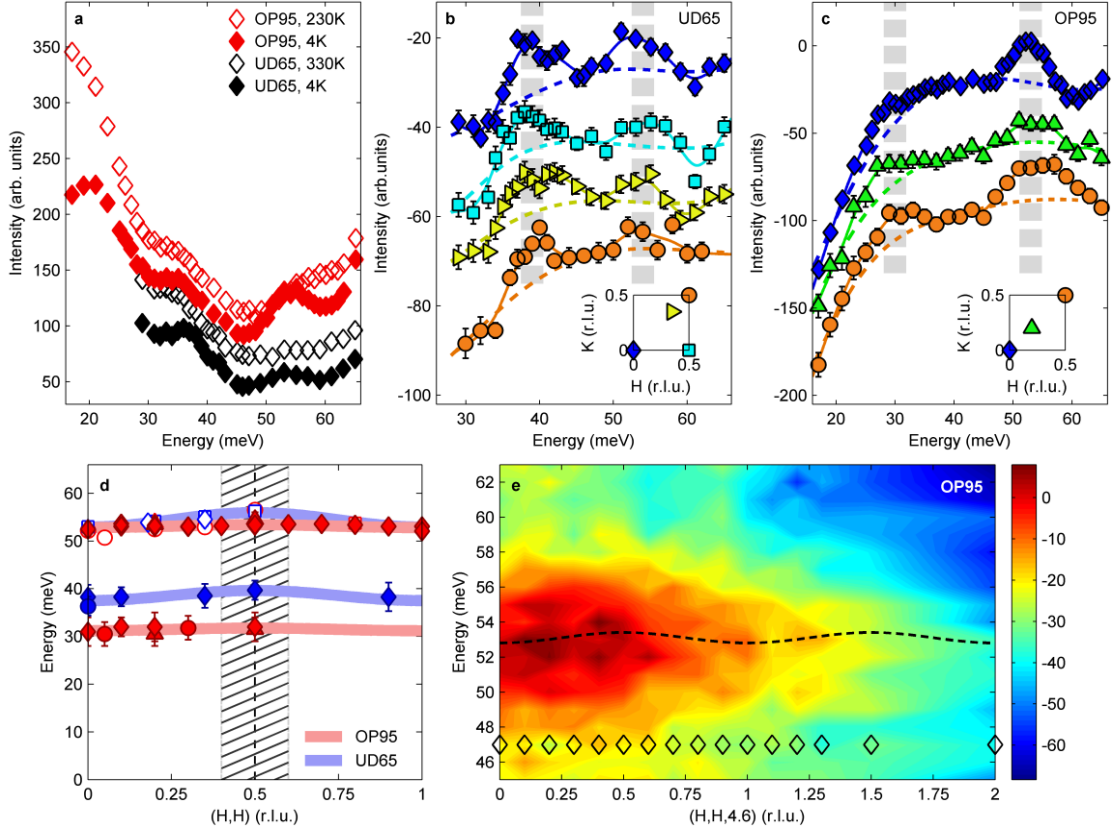

**Figure 1 | Observation of two excitation branches. a**, Unpolarized inelastic neutron scattering data at $\mathbf{Q} = (0, 0, 4.6)$. The high-energy (~53 meV) magnetic excitation reported in ref. 24 is evident from the 4 K data; the low-energy excitation is difficult to discern from the raw spectra due to phonons nearby. **b**, Intensity difference between 4 K and 330 K (top three) and between 4 K and 300 K (bottom) for UD65 ($T^* \approx 330$ K[24]) at $\mathbf{Q} = (0, 0, 4.6)$, (0.5, 0, 4.6), (0.35, 0.35, 4.6), and (0.5, 0.5, 4.4), from top to bottom. The bottom data set was obtained with better energy resolution (~ 4 meV (FWHM) at $\omega = 40$ meV, compared to ~ 6 meV for the rest). **c**, Intensity difference between 4 K and 230 K (top) and between 4 K and 200 K (bottom two, measured on a different spectrometer and rescaled for comparison) for OP95 ($T^* \approx 210$ K[24]) at $\mathbf{Q} = (0, 0, 4.6)$, (0.2, 0.2, 4.4) and (0.5, 0.5, 4.4), from top to bottom. In **b** and **c**, the solid lines are guides to the eye, and the data are offset for clarity (top data sets are

without offset). The magnetic signal is superposed on a baseline (dashed lines) that is more negative at lower energies due to the stronger increase of phonon scattering toward high temperatures. The insets indicate the measured 2D momentum positions. **d**, Dispersion along [*H,H*] of the two excitations observed at the two doping levels. Empty symbols are data reported in ref. 24. Different symbols indicate on which spectrometers the measurements were performed (diamond: IN8; circle: IN20; square: PUMA; triangle: 2T). Hatched area indicates where antiferromagnetic spin fluctuations are expected. Error bars in **a-c** represent statistical uncertainty (1 s.d.), and in **d** the confidence range for the estimated energies. **e**, Color representation of intensity difference between 4 K and 230 K for the high-energy excitation (dashed line) of OP95. Diamond symbols indicate momentum positions where energy scans were performed.

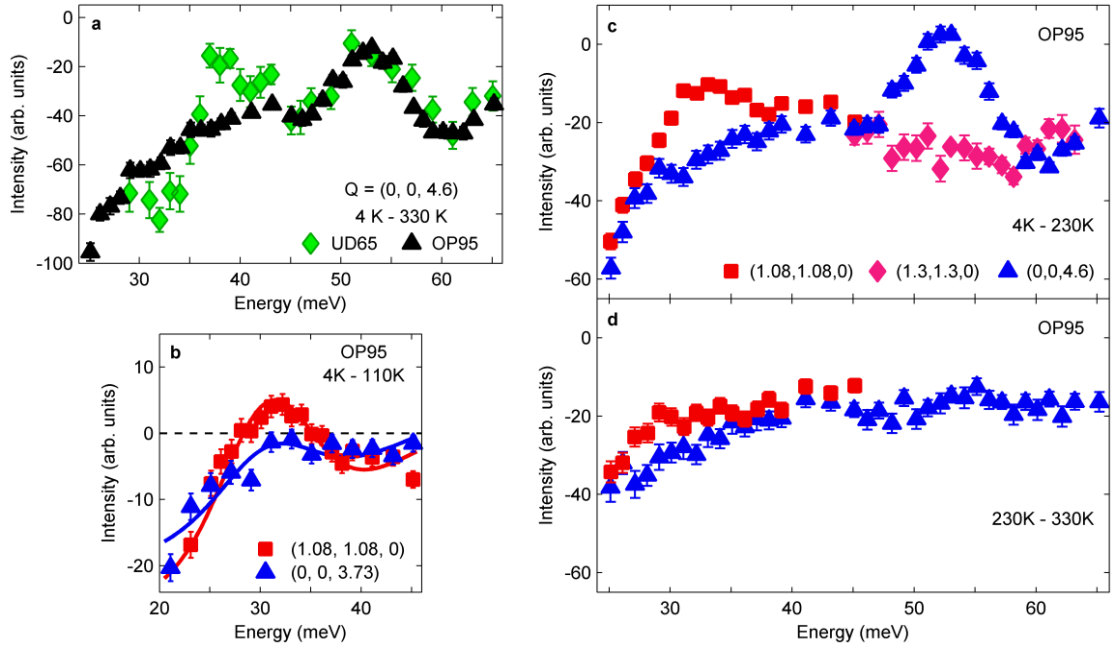

**Figure 2 | Doping and momentum dependence of intensity. a**, Doping dependence of low-energy excitation revealed by net intensity ('4 K - 330 K') for both samples. The measurements were performed on the same spectrometer with similar configuration. The UD65 data are rescaled to the high-energy (~ 53 meV) signal of the OP95 data. **b**, Intensity difference between 4 K and 110 K measured with $\mathbf{Q}\|c$ and $\mathbf{Q}\|ab$ for the low-energy excitation in sample OP95. The magnitudes of the two momenta (1.08, 1.08, 0) and (0, 0, 3.73) are identical. By using a low reference temperature of 110 K, the increase of the magnetic signal at ~ 31 meV toward 4 K overcomes the decrease in phonon scattering, leading to a net intensity increase near 31 meV for $\mathbf{Q}$ = (1.08, 1.08, 0). **c**, Intensity difference between 4 K and 230 K from energy scans for OP95. In contrast to **b**, the low-energy excitation at $\mathbf{Q}$ = (0, 0, 4.6) is difficult to discern from these data because of the baseline slope due to phonons. The magnitudes of the momenta (1.3, 1.3, 0) and (0, 0, 4.6) are identical. **d**, Intensity difference between 230 K and 330 K for OP95. The $\mathbf{Q}$ = (1.08, 1.08, 0) data lie above

**Q** = (0, 0, 4.6) because the baseline due to phonons is less negative at smaller $Q$, and therefore the difference is not necessarily a low-energy magnetic signal. Error bars represent statistical uncertainty (1 s.d.).

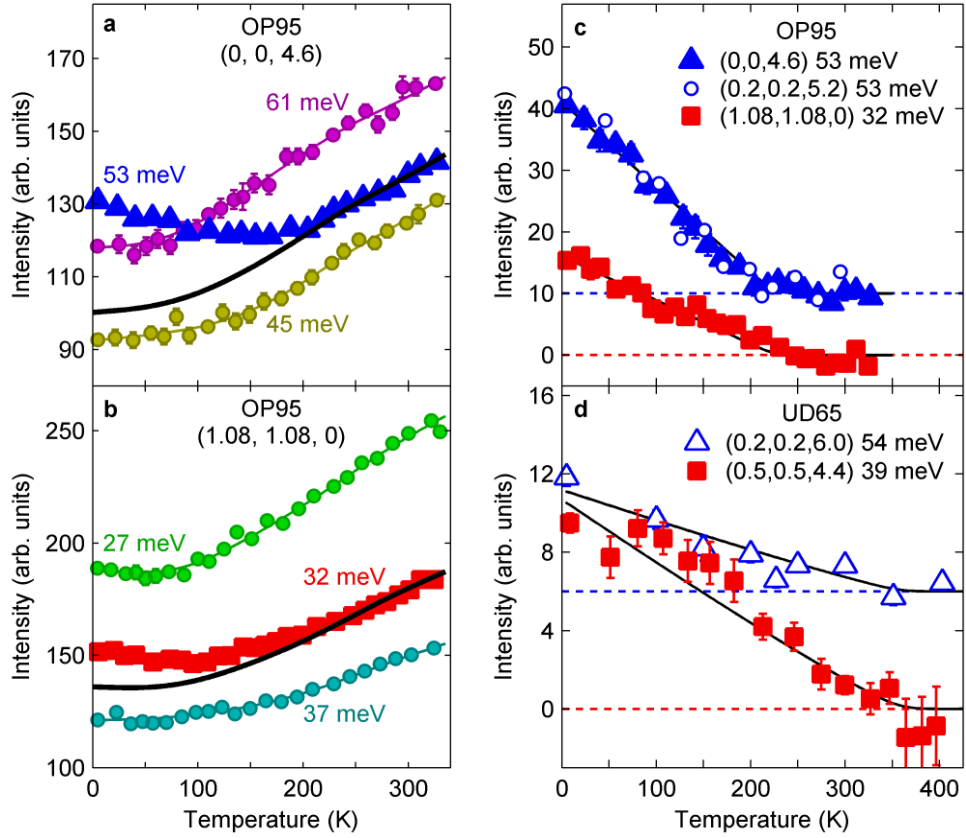

**Figure 3 | Temperture dependence. a,b,** Temperature dependence of intensities at the indicated energies and momenta for OP95. **c,** The temperature dependence of the pseudogap excitations is obtained by subtracting from the data at 53 meV and 32 meV the background intensities, which are estimated as the average of the smoothed temperature dependences (thin lines in **a,b**) at higher and lower energy with a vertical offset (black lines in **a,b**). **d,** The same method is used to obtain the temperature dependence of the pseudogap excitations for UD65. Data in **c** and **d** are offset for clarity (dashed lines), and the empty symbols are data reported in ref. 24. Solid lines are guides to the eye. Error bars represent statistical uncertainty (1 s.d.).

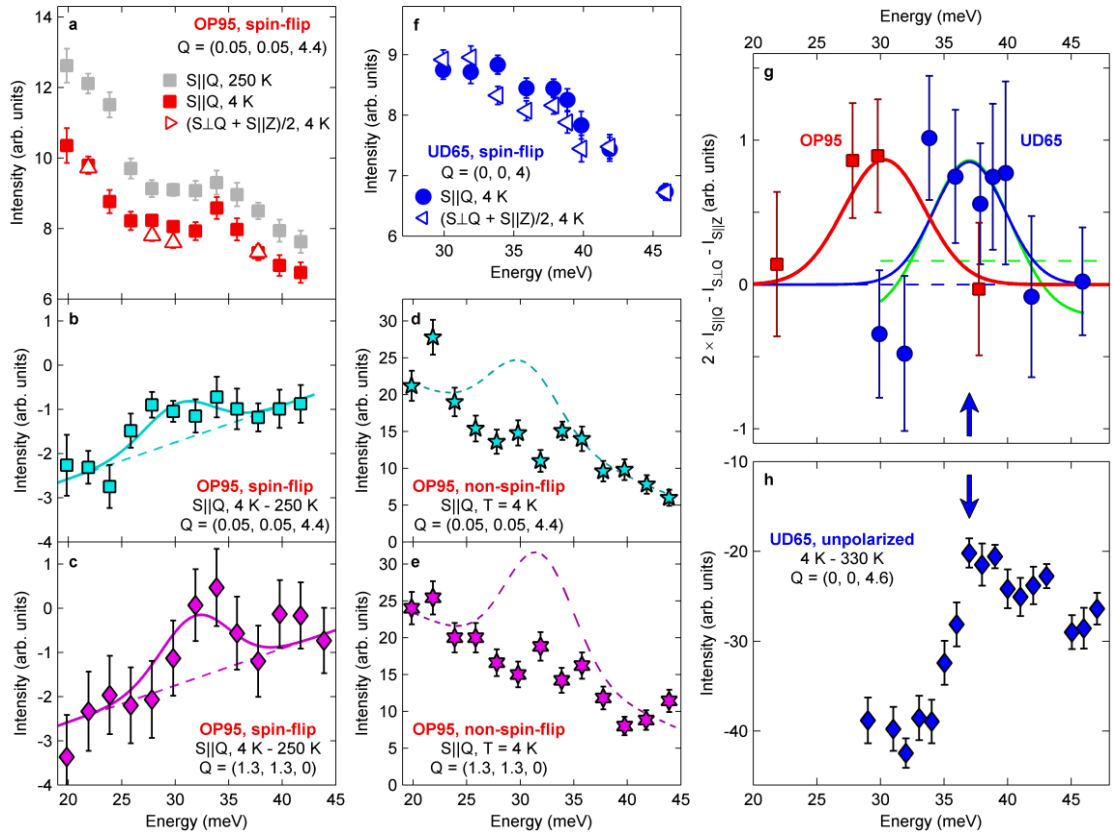

**Figure 4 | Magnetic origin verified by spin-polarized measurements. a**, Spin-flip spectra at $\mathbf{Q} = (0.05, 0.05, 4.4)$ for OP95. Filled symbols are measured with the initial neutron spin polarization ($\mathbf{S}$) parallel to $\mathbf{Q}$, a geometry in which all magnetic fluctuations are probed. Empty symbols are the average of intensities measured with $\mathbf{S}$ in the horizontal scattering plane but perpendicular to $\mathbf{Q}$ ($\mathbf{S} \perp \mathbf{Q}$) and with $\mathbf{S}$ vertical ($\mathbf{S} \| \mathbf{Z}$), which measures only half of the total magnetic signal (Supplementary Fig. S5 shows that $I_{\mathbf{S} \perp \mathbf{Q}}$ and $I_{\mathbf{S} \| \mathbf{Z}}$ are the same within the error, consistent with the system's tetragonal symmetry). **b,c**, Intensity difference between 4 K and 250 K for OP95 measured in the $\mathbf{S} \| \mathbf{Q}$ spin-flip geometry at $\mathbf{Q} = (0.05, 0.05, 4.4)$ and $(1.3, 1.3, 0)$, respectively. Solid lines are Gaussian fits assuming a common width and baseline. **d,e.** Non-spin-flip intensity at 4 K for sample OP95 at $\mathbf{Q} = (0.05, 0.05, 4.4)$ and

(1.3, 1.3, 0). Dotted lines illustrate the size of non-spin-flip nuclear (phonon) signal that would be required to produce the peaks in **b,c** via polarization leakage given the instrumental flipping ratio of ~ 10. **f**, Spin-flip data at $\mathbf{Q} = (0, 0, 4)$ for UD65. **g**, Magnetic signal extracted from polarization analysis of the 4 K data in (a&f). Solid blue line is the best Gaussian fit to the data for UD65 assuming zero offset. Solid and dashed green lines are best Gaussian and constant fits which allow for a non-zero offset. Red line is adapted from the fit in **b** without the linear baseline. A statistical analysis of the data is presented in the SI **h**, Unpolarized neutron data for UD65 adapted from Fig. 1b to directly demonstrate that the magnetic signal in **g** occurs at the peak position of the unpolarized result for closely similar values of $\mathbf{Q}$. Error bars represent statistical uncertainty (1 s.d.).

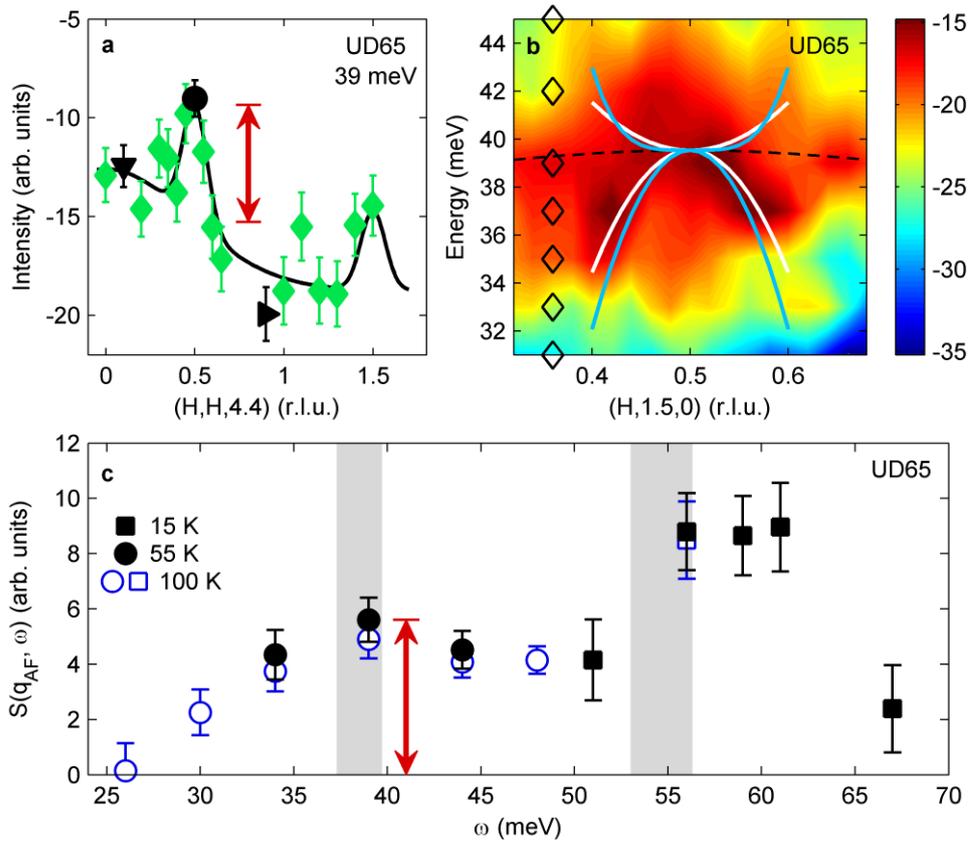

**Figure 5 | Mixing between pseudogap excitations and antiferromagnetic spin fluctuations. a**, Intensity difference between 4 K and 300 K at $\omega = 39$ meV for UD65. Solid line is a guide to the eye, which comprises a decrease of the $q = 0$ signal toward large $Q$ and additional intensities at $\mathbf{q}_{AF}$. Black symbols are data shared in common with the energy scans in Supplementary Fig. S6a. Error bars represent statistical uncertainty (1 s.d.). **b**, Color representation of intensity difference between 4 K and 330 K for UD65. Data are smoothed along the horizontal axis for improved visual inspection (Supplementary Fig. S7). Diamond symbols indicate energy positions of the underlying momentum scans. Solid lines illustrate the typical 'hourglass' dispersion of the spin fluctuations near $\mathbf{q}_{AF}$ in YBCO, adapted from ref. 37 (blue) and ref. 46 (white). Dashed line indicates the dispersion of the pseudogap excitation. **c**,

Amplitude of antiferromagnetic response (peak at (0.5, 0.5) as indicated by the arrows) for UD65 as a function of energy ($\omega$), measured on the spectrometers 2T (circles) and PUMA (squares). Error bars represent fit uncertainty for signal amplitudes observed in individual scans (Supplementary Fig. S6b,c). Shaded areas illustrate the full dispersion of the pseudogap excitations (maximal dispersion energies reached at $\mathbf{q}_{AF}$).

# Supplementary Information

This document includes supplementary text, Figures S1-S7, and references.

## Supplementary text

Our spin-polarized measurements (Figs. 4, S4b, S5) were carried out using the thermal neutron beam on the IN20 spectrometer at the Institute Laue Langevin, France, currently the world's best instrument of its kind. Heusler alloy crystals were used as monochromator and analyzer, which select the initial and final neutron energy as well as the spin polarization of the neutrons. The polarization of the neutron beam around the sample position was maintained by CryoPAD[47]. This device provides high stability and reproducibility of the neutron spin polarization among all of our measurement conditions.

Spin-unpolarized measurements were performed on spectrometer 2T (bottom two data sets in Fig. 1c, Fig. S6b) at the Laboratoire Léon Brillouin, France, on spectrometer PUMA (Fig. S6c) at the Forschungs-Neutronenquelle Heinz Maier-Leibnitz, Germany, and on spectrometer IN8 (all other figures) at the Institute Laue Langevin, France. Pyrolytic graphite analyzers and pyrolytic graphite (or copper, which provides the better energy resolution in Figs. 5a and S6a, and the bottom data set in Fig. 1b) monochromators were used in the unpolarized measurements. Pyrolytic graphite filters were used in all measurements to suppress harmonic scattering on the analyzer, where the final neutron energy was fixed at either 30.5 meV or 35 meV. The typical energy resolution was ~ 5 meV (FWHM) in the 30-40 meV energy transfer range and ~ 8 meV (FWHM) in the 50-60 meV energy transfer range.

In all figures except for Fig. S4, the intensity units are counts per ~ 12 seconds on spectrometers IN8 (unpolarized) and ~ 80 seconds on IN20 (polarized). Data obtained on other spectrometers (unpolarized) are rescaled to allow a common vertical scale.

In a spin-polarized neutron scattering experiment, it is convenient to define the coordinate system for the spin polarization ($\mathbf{S}$) relative to the scattering geometry, *i.e.*, with the three principal axes $\mathbf{S}\|\mathbf{Q}$, $\mathbf{S}\perp\mathbf{Q}$, and $\mathbf{S}\|\mathbf{Z}$, where $\mathbf{Q}$ is the momentum transfer. The polarizations $\mathbf{S}\|\mathbf{Q}$ and $\mathbf{S}\perp\mathbf{Q}$ lie in the horizontal scattering plane, whereas $\mathbf{S}\|\mathbf{Z}$ is vertical. In the absence of chiral magnetic correlations, the measured spin-flip scattering intensities in the three spin-polarization geometries correspond to[48]:

$I_{\mathbf{S}\|\mathbf{Q}} = 2/3 \times N_{\text{inc,spin}} + M_{\perp\mathbf{Q}} + M_{\|Z} + \text{BG},$

$I_{\mathbf{S}\perp\mathbf{Q}} = 2/3 \times N_{\text{inc,spin}} + M_{\|Z} + \text{BG},$

$I_{\mathbf{S}\|Z} = 2/3 \times N_{\text{inc,spin}} + M_{\perp\mathbf{Q}} + \text{BG},$

where $N_{\text{inc,spin}}$ is the nuclear spin incoherent cross-section, $M$ is the magnetic cross-section that corresponds to magnetic fluctuations along the indicated direction, and BG is the background contribution. In our spin-polarized measurements, $\mathbf{Q} = (H, H, L)$ lies in the horizontal scattering plane. For $L \neq 0$ and $H \approx 0$, as is the case in Figs. 4 and S5, "$\perp\mathbf{Q}$" and "$\|\mathbf{Z}$" correspond to the crystallographic [110] and [1-10] directions, respectively, which are equivalent given the tetragonal symmetry of Hg1201. The magnetic signals observable in the $\mathbf{S}\perp\mathbf{Q}$ and $\mathbf{S}\|\mathbf{Z}$ geometries should therefore be equal to half of that in the $\mathbf{S}\|\mathbf{Q}$ geometry. The data in Fig. S5 are consistent with this. Regardless of this simplification by symmetry, it is always true that $2 \times I_{\mathbf{S}\|\mathbf{Q}} - I_{\mathbf{S}\perp\mathbf{Q}} - I_{\mathbf{S}\|Z} = M_{\perp\mathbf{Q}} + M_{\|Z}$, which we use to extract the magnetic signal in

Fig. 4g.

The combined unpolarized and polarized neutron scattering measurements therefore demonstrate that the low-energy excitation around 32 meV is of predominant magnetic origin: given that the data in Fig. 4g are randomly distributed about 'true' values (with Gaussian probability and standard deviation given by the error bars), the likelihood for the 'peak scenario' (solid blue) curve centered at the same energy as the unpolarized result (Fig. 4h) to describe the UD65 data is a factor of 180 higher than the zero-level (dashed blue line). Even if a systematic error resulted in a non-zero baseline in the absence of any magnetic signal, the 'peak scenario' (solid green curve) is still better than the 'no peak scenario' (dashed green line) by a factor of 25.

**Supplementary Figures**

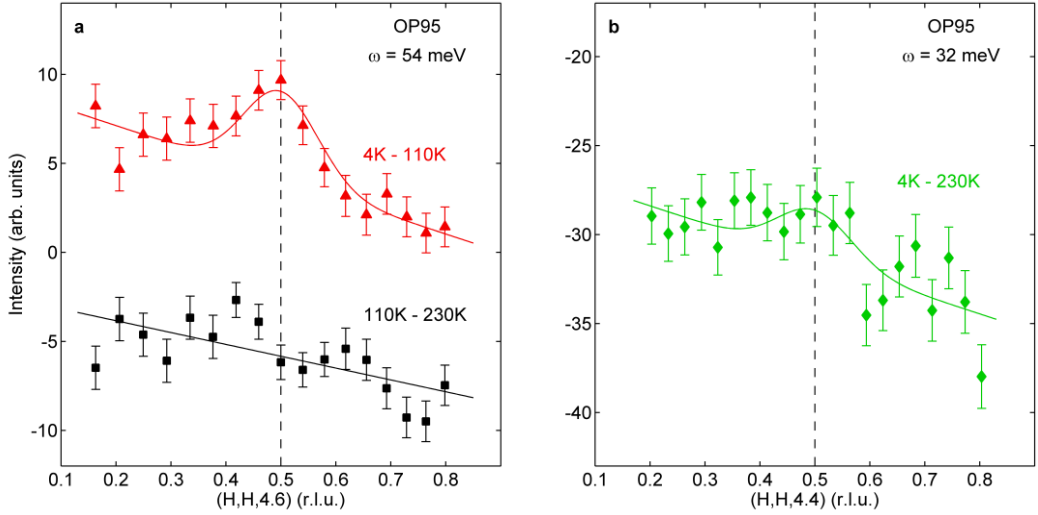

**Figure S1. a**, Momentum scans at 54 meV for OP95 show a magnetic resonance peak at $\mathbf{q}_{AF}$ in the difference between 4 K ($< T_c$) and 110 K ($> T_c$), consistent with previous work[30]. Part of the slope is due to the increasing intensity of the underlying high-energy magnetic pseudogap excitation toward $H = 0$. **b**, A local intensity maximum at $H = 0.5$ is also observed for the low-energy magnetic pseudogap excitation in OP95, but the relatively weak signal does not allow for a detailed determination of whether it exhibits a resonance-like intensity change across $T_c$. Local intensity maxima at $\mathbf{q}_{AF}$ are also observed at the same energies as the two pseudogap excitations in UD65, but the signal sets in already at $T^*$ and exhibits no distinct change across $T_c$ (Fig. 3d), similar to the normal-state response at $\mathbf{q}_{AF}$ in underdoped YBCO[49,50]. Although a precise comparison of the signal amplitudes at $\mathbf{q}_{AF}$ between YBCO and Hg1201 is complicated by the different crystal structures and the strong neutron absorption of Hg, having measured both systems ourselves, we assess that the signals (per planar Cu site) are comparable in strength within a factor of two.

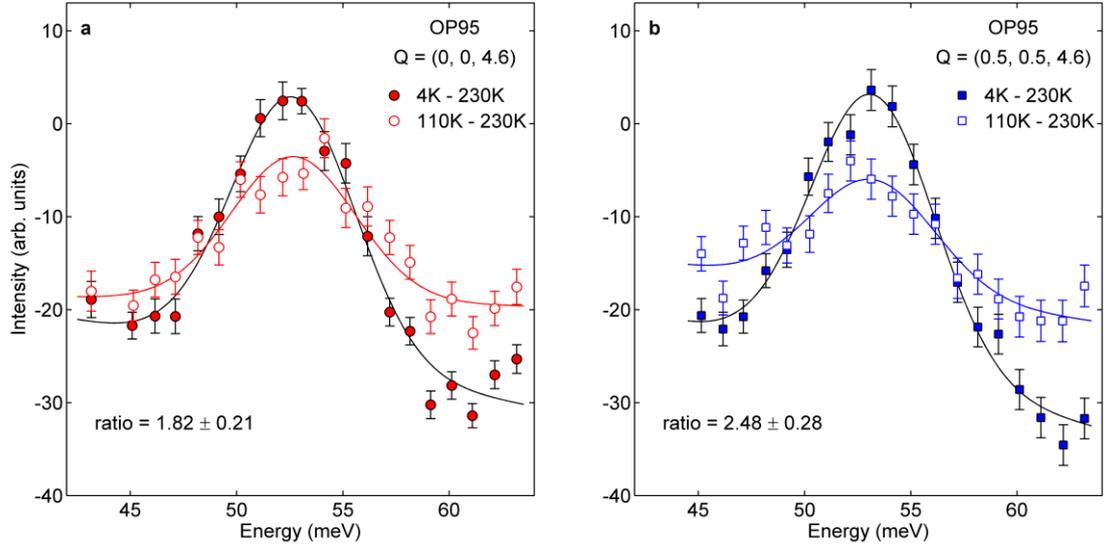

**Figure S2.** Comparison of the intensity of the high-energy pseudogap excitation in sample OP95 between 4 K and 110 K (> $T_c$) at the 2D Brillouin zone center (**a**) and at $\mathbf{q}_{AF}$ (**b**). Intensity measured at 230 K (> $T^*$) has been subtracted from the data. The higher intensity ratio $I$(4 K) / $I$(110 K) at $\mathbf{Q}$ = (0.5, 0.5, 4.6) is due to the additional presence of the magnetic resonance below $T_c$. Based on these data, we conclude that the intensity ratio (at 4 K) between the resonance and the pseudogap excitation at $\mathbf{Q}$ = (0.5, 0.5, 4.6) and 53 meV is approximately 1:2.7.

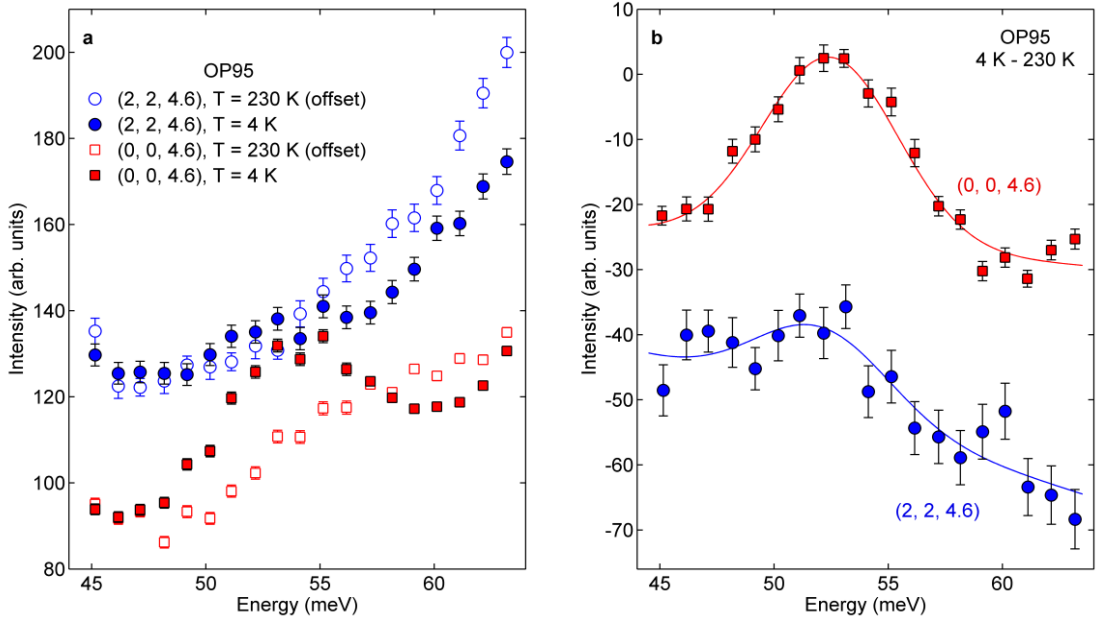

**Figure S3**. Intensity of the high-energy excitation in sample OP95 at both zero and large in-plane momentum transfer. In **a**, the data for $T = 230$ K are vertically offset to allow better comparison with the 4 K data. Given the crystal structure of Hg1201, $H = K = 0$ and $H = K = 2$ are equivalent concerning the relative phases of different nuclei's contributions to the phonon structure factor. Since the Debye-Waller factor is unimportant at low temperature, any phonon-related effect ought to be stronger (or, at least, not weaker) at $H = K = 2$ than at $H = K = 0$. Therefore, the stronger effect observed at $H = K = 0$ cannot be interpreted as a phonon anomaly.

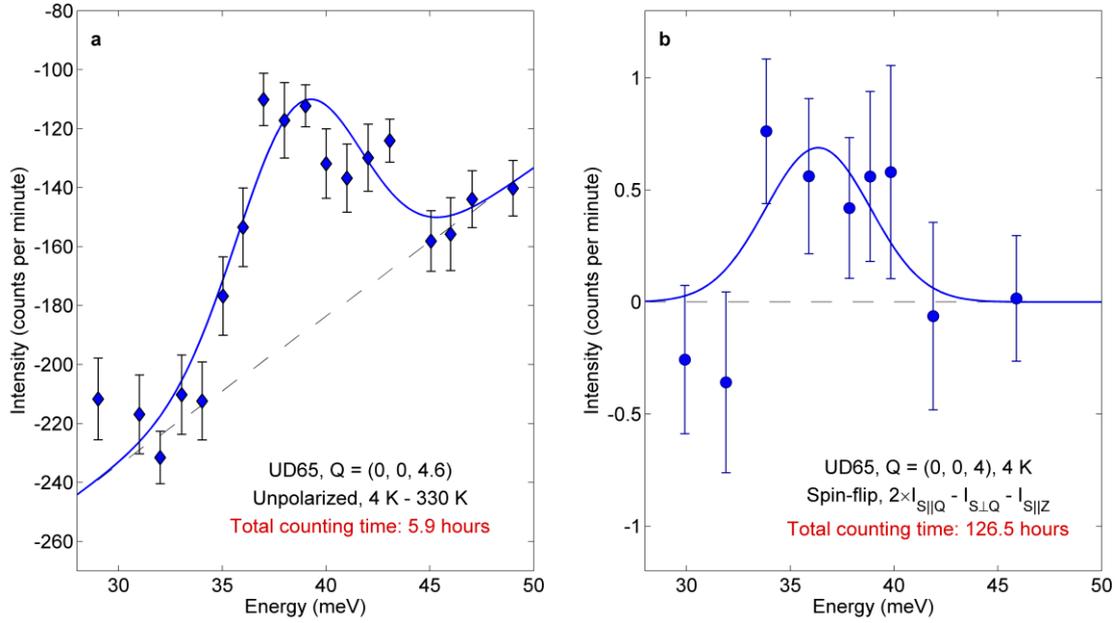

**Figure S4**. Comparison between measurements of the low-energy excitation in sample UD65 performed on spin-unpolarized (IN8 at ILL; from Fig. 1b) and polarized (IN20 at ILL; from Fig. 4g) instruments. Solid lines are best fits to the data with a Gaussian profile. Error bars represent statistical uncertainty (1 s.d.). As can be seen from the total counting time, the unpolarized measurement benefits from a much higher neutron flux, but the subtraction of background measured at high temperature may introduce additional non-statistical errors. The spin-polarized measurement allows for a more reliable determination of the genuine magnetic signal, but the low neutron flux and the algebra involved in the polarization analysis require a long counting time, which prohibits high statistical precision. These limitations in the measurement accuracy account for the discrepancy of 2 meV between the fitted peak positions in **a** and **b**.

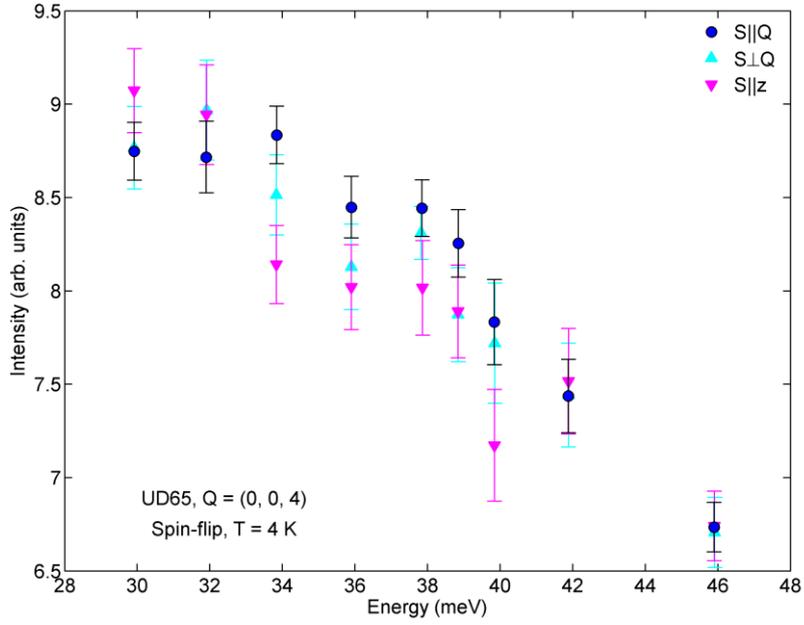

**Figure S5**. Raw data of spin-flip spectra for sample UD65 measured at **Q** = (0, 0, 4) and *T* = 4 K in different spin-polarization geometries. The intensities in the **S**⊥**Q** and **S**∥**Z** geometries are consistent with each other within the error, and they are systematically lower than the intensity in the **S**∥**Q** geometry in the range 34 − 40 meV, as expected in the presence of a genuine magnetic signal centered at ∼ 37 meV. Error bars represent statistical uncertainty (1 s.d.).

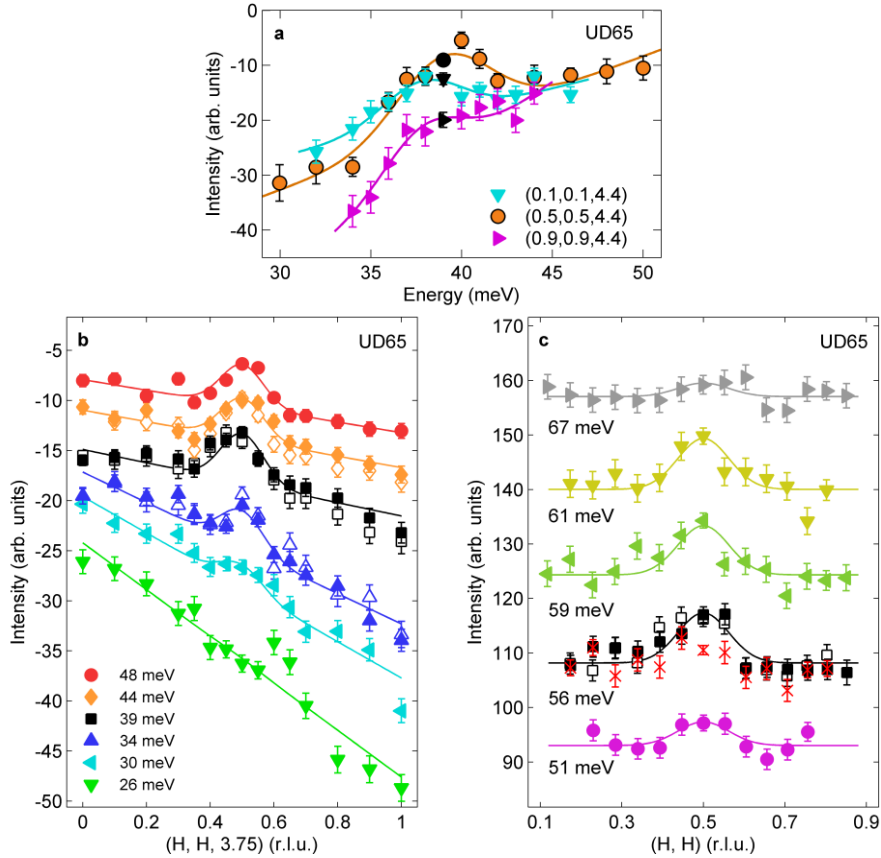

**Figure S6**. **a**, Intensity difference between 4 K and 300 K for UD65 using the same high-energy-resolution configuration as for the bottom data set in Fig. 1b. Solid lines are Gaussian fits assuming a linear slope. Black symbols are data shared in common with Fig. 5a. **b**, Intensity difference between low temperatures and 370 K for UD65. Filled symbols: $T = 100$ K; empty symbols: $T = 60$ K. Solid lines are Gaussian fits to the 100 K data. **c**, Raw intensity of rocking (constant-$Q$) scans in which the value of $L$ varies between 4.4 and 5.2. Filled symbols: $T = 15$ K; empty squares: $T = 100$ K for 56 meV; red crosses: $T = 395$ K for 56 meV. Lines are Gaussian fits to the 15 K data. Except for 48 and 51 meV, the data in b and c are offset for clarity. Note that the scans in b and c are along $[H, H]$, which is not as optimized for observing the "hourglass" dispersion as the $[H, 0]$ scan direction in Fig. 5b. Error bars represent statistical uncertainty (1 s.d.).

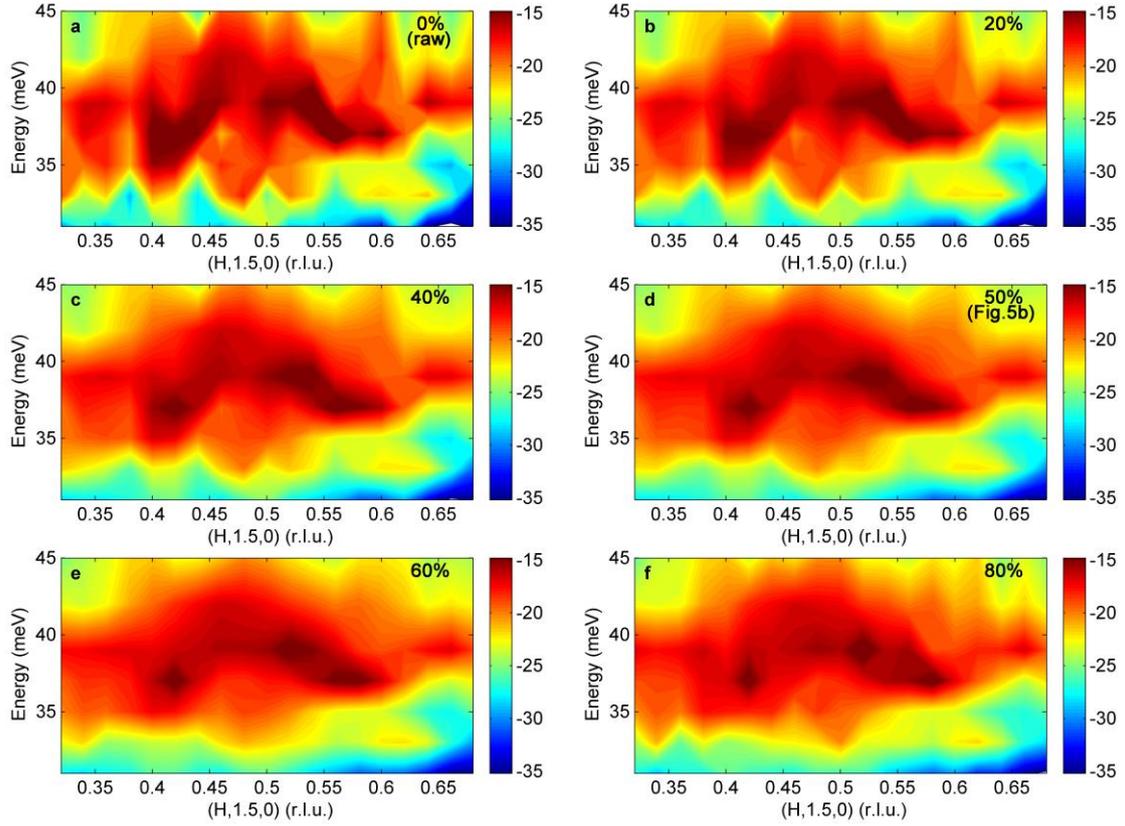

**Figure S7.** Color representation of intensity difference between 4 K and 330 K for the low-energy excitation in UD65 near (0.5,1.5,0) with different degrees of smoothing. Panel **d** is the same as Fig. 5b. The smoothing is performed along the horizontal axis by taking the weighted average of the intensity at each $H$ with the values of the two adjacent data points at the same energy. The total weight on the side points is indicated (*e.g.*, "60%" means "*left*\*30% + *center*\*40% + *right*\*30%"). The concave dispersion near $H = 0.5$ is robust regardless of the degree of smoothing.

**Supplementary References**